# Electrostatic Affinities and Binding Kinetics among $\alpha_2$-I Integrin Domains, Divalent Cations, and 21-mer Collagen Fragment

Patrizio Ansalone,
Istituto Nazionale di Ricerca Metrologica (INRIM), Strada delle Cacce 91, Torino 10135, Italy.
p.ansalone@inrim.it

**Background:** In the interaction between $\alpha_2$-I integrin domain and collagen, there is a competition between the geometric constraint on possible reciprocal orientations and long-range electrostatic force. Published experimental results have shown that the $k_{on}$ rate constant, for the 42-mer-collagen fragment and the $\alpha_2$-I integrin domain complex, ranges in order of magnitude from $10^4$-$10^5$ $M^{-1}s^{-1}$. This is the lower bound of the interval determined by the diffusion-limited regime and the orientational constraint.

**Results**: The electrostatic affinity between collagen and the $\alpha_2$-I integrin domain can be expressed in terms of the Pearson correlation coefficient between the desolvation potential of the ligand and the interaction potential of the receptor, and is a measure of electrostatic complementarity. Simulations of atomistic Brownian dynamics were performed to probe and characterize the association process. It was found that association values are compatible with published experimental data.

**Conclusions**: In the bound state, the integrin-collagen structures clearly reveal electrostatic complementarity in a specific reaction patch. In the analysis, electrostatic and steric effects of divalent cations ($Mg^{2+}$, $Co^{2+}$, $Mn^{2+}$ and $Ca^{2+}$) are discussed, which are located in the metal-ion-dependent site of the $\alpha_2$-I domain. Simulations of Brownian dynamics in the diffusion-limited regime confirm that this approach faithfully describes the binding kinetics of the $\alpha_2$-I integrin domain and the 21-mer collagen in presence of $Mg^{2+}$, $Co^{2+}$ and $Mn^{2+}$ according to published experimental data. The Poisson Boltzmann model requires additional modification to provide a reasonable description of divalent cations. Finally the geometrical parameters and flexibility of the binding pocket in the $\alpha_2$-I integrin collagen complex should be update from quantum chemical calculations.

## Background

This work provides a mechanistic insight into the binding process between collagen and the $\alpha_2$-I integrin actually this process is not fully understood and its comprehension and it has relevant consequence in the field of regenerative medicine. Moreover, deeper knowledge concerning the binding mechanism at the macromolecular level would help to understand interactions between cells and surrounding biological tissues. For these reasons, integrin-mediated cellular interactions with the environment [1] is a important research area. Of particular interest is the cellular adhesion that occurs on "scaffolds," which are artificial organic or inorganic engineered extracellular matrices able to support tissues. The most experimentally studied, collagen-binding integrins are $\alpha_1\beta_1$ and $\alpha_2\beta_1$. Both bind with high affinity to type I and IV triple-helical peptides [2]. Integrin receptors are the primary cell adhesion proteins that mediate extracellular matrix and cell-cell interactions [3][4]. As suggested previously, divalent cations such as $Mg^{2+}$, $Co^{2+}$, $Mn^{2+}$, and $Ca^{2+}$ [5][6] are key factors in promoting or inhibiting the binding process. Protein candidates for the analysis are the $\alpha_2$-I domain of the $\alpha_2\beta_1$ integrin transmembrane metalloprotein and a fragment of the 21-mer collagen. The analysis focuses on the complex defined by pdb code 1DZI. The building blocks are specific I-domain integrin α subunits (*i.e.,* α2-I) in the active state conformation, and a small fragment of triple-helical-collagen peptide type-I [7][8][9][10]. They are available on the Research Collaboratory for Structural Bioinformatics (RCSB) pdb archive [11].

## Methods

### Numerical model

I focus on the electrostatic affinity in the binding of *receptor* and *ligand* proteins in an aqueous ionic environment. The problem is modeled by considering the linearized approximation of the Poisson-Boltzmann equation (LPBE) that is solved by using a boundary element method (BEM) approach. The domain composed of an external open boundary region $\Omega^s \subset R^3$, representing the solvent, and a molecular region $\Omega^m = \Omega_R^m \cup \Omega_L^m$, where $\Omega_R^m$ and $\Omega_L^m$ are the receptor and ligand molecular regions, respectively. The molecular surface $\Sigma_{\Omega^m}$ is defined as the boundary of the union of the molecular regions, *i.e.,* $\Sigma_{\Omega^m} = \partial(\Omega_R^m \cup \Omega_L^m)$. The solvent region $\Omega^s$ is described as a continuum dielectric medium, where ionic species satisfy the Boltzmann distribution. In this region, characterized by a uniform relative dielectric constant $\varepsilon_s$, the LPBE assumes the form:

$$\Delta\Phi^s - \kappa^2\Phi^s = 0 \quad \text{in} \quad \Omega^s, \tag{1}$$

Equation 1 [12] is valid if the electrostatic energies of the mobile ions in the solvent are smaller than their thermal energies, and if $|e_c\Phi(\mathbf{x})| \ll k_B T$. The electrostatic potential $\Phi^s$ is the unknown in the solvent region. The parameter $\kappa$ is the Debye-Hückel electric field screening parameter (*i.e.,* the reciprocal of the Debye length), and is given by $\kappa = \sqrt{2N_A e_c^2 I_s (k_B T \varepsilon_s \varepsilon_0)^{-1}}$, where $I_s$ is the solvent ionic strength, $e_c$ is the electron charge, $N_A$ is the Avogadro's constant, $T$ is the absolute temperature, $k_B$ is Boltzmann's constant, $\varepsilon_0$ is the vacuum dielectric constant, and $\varepsilon$ is the dielectric constant in the solvent region. In the molecular regions $\Omega_L^m$ and $\Omega_R^m$, the LPBE equation reduces to the Poisson equation:

$$\Delta\Phi^m = -(\varepsilon_m \varepsilon_0)^{-1} \sum_{i=1}^{N_q} q_i \delta^{(3)}(\mathbf{x} - \mathbf{x}_i), \quad \text{in} \quad \Omega_R^m \cup \Omega_L^m \tag{2}$$

In Eq. 2, $\Phi^m$ is the electrostatic potential inside the molecular regions, $x_i|_{i=1\cdots N_q}$ is the location of $N_q$ point charges, $q_i|_{i=1\cdots N_q}$ represents their strengths, the relative dielectric constant $\varepsilon_m$ of the molecules is usually in the range 1-4, and $\delta^{(3)}(\mathbf{x} - \mathbf{x}_i)$ is the Dirac distribution. Several numerical methods exist for solving the displacement of the electrostatic field induced by protein charge distributions, including the [13] finite difference method [14], the finite element method [15][16], the element-free Galerkin method [17], and BEM [18]. Here, the LPBE is solved introducing a BEM formulation.

### Residual potential and electrostatic affinity

The electrostatic affinity between the $\alpha_2\beta_1$ integrin and a 21-mer collagen is investigated on the basis of residual potential and complementarity [19], assuming that one protein acts as a ligand and the other a receptor. The residual potential $\Phi(\mathbf{x})$ is computed as the sum of two terms. The first is the interaction potential $\Phi_R^i(\mathbf{x})$, which is defined as the electrostatic potential of the bound state due to the receptor charges. The second is the desolvation potential $\Phi_L^d(\mathbf{x})$, which is the difference between the electrostatic potential due to the charges of the ligand on the bound state and the electrostatic potential due to the charges of the ligand in the unbounded state, $\Phi(\mathbf{x}) = \Phi_R^i(\mathbf{x}) + \Phi_L^d(\mathbf{x})$. Therefore, the residual potential can be obtained by solving LPBE three times for the following cases:

a) Domain containing the dielectric region $\Omega_R^m \cup \Omega_L^m$, with only receptor charges.

b) Domain containing the dielectric region $\Omega_R^m \cup \Omega_L^m$, with only ligand charges.

c) Domain containing the dielectric region $\Omega_L^m$, with the ligand charges.

In cases a) and b), the problem is solved by discretizing the molecular surface $\Omega_R^m \cup \Omega_L^m$. In case c), only the molecular surface $\Omega_L^m$ is meshed, introducing a grid of N nodes. The residual potential provides direct information on the electrostatic affinity, and, in the presence of high electrostatic complementarity, it approaches zero in the ligand volume and on its molecular surface [20],[21]. In this case, the interaction potential must be almost equal in magnitude, but opposite in sign, to the desolvation potential. This effect is known as the "complementarity principle." The extent of the electrostatic complementarity of the ligand for its receptor can be determined by the Pearson correlation coefficient $r_{\Phi_R^i \Phi_L^d} = \dfrac{\text{Cov}(\Phi_R^i, \Phi_L^d)}{\sqrt{\text{Var}(\Phi_R^i)}\sqrt{\text{Var}(\Phi_L^d)}}$ [22].

**Binding kinetics**

The concept of ligand-receptor interaction is considered here to study site-specific binding kinetics between the $\alpha_2$-I domain and the 21-mer collagen fragment [23]. They should exhibit different kinetics depending on low, weak, or high reciprocal affinity in an ionic aqueous environment. The formation of a ligand-receptor encounter complex is represented by the three-state equilibrium binding reaction:

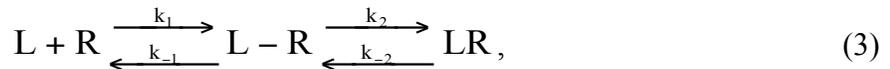

$$ L + R \underset{k_{-1}}{\overset{k_1}{\rightleftarrows}} L-R \underset{k_{-2}}{\overset{k_2}{\rightleftarrows}} LR, \qquad (3) $$

where L and R are the free ligand and receptor proteins, respectively, and L-R is the ensemble of allowed orientations that precedes the LR complex. During the first step, the electrostatic interaction exerts its role in the initial encounter. The second step represents bond formation due to stereospecific association. Then, starting from the results in [24], the "on" rate constant $k_{on} = \dfrac{k_1 \cdot k_2}{k_{-1} + k_2}$ is quantified for different metal ions in the diffusion-limited regime based on simulations of Brownian dynamics.

**Results**

**Electrostatic affinity of the $\alpha_2$-I domain and the 21-mer collagen fragment with and without $Mg^{2+}$, $Co^{2+}$, $Mn^{2+}$ and $Ca^{2+}$**

Here, attention is focused on the electrostatic interaction between a triple helical collagen fragment and the $\alpha_2$-I domain of the integrin $\alpha_2\beta_1$. Specifically, the $\alpha_2$-I domain in the open conformation of the $\alpha_2\beta_1$ integrin is considered; this corresponds to the high-affinity state [7]. To determine the influence of metal ions located in the metal-ion-dependent site (MIDAS) on the electrostatic binding affinity of the $\alpha_2$-I domain, the binding affinity is evaluated by computing the residual potential on the ligand molecular surface. This is performed in accordance with the definition in [19], [20], and [21], and by comparing the different results in terms of the Pearson correlation coefficient. In the analysis, it is assumed that the $\alpha_2$-I domain could be bonded to a metal ion as the ligand (L), and the collagen as the receptor (R). The simulations are performed based on the two molecular structures extracted from the 1DZI complex. The final model includes the $\alpha_2$-I domain, a metal ion, and three strands of the collagen-like peptide that binds to the $\alpha_2$-I domain in a unique orientation with staggered strands [9]. The reconstruction of missing atoms, hydrogen optimization, and the generation of input parameters (*i.e.*, atomic charges and radii) for the Poisson-Boltzmann simulations are performed with the PDB2PQR pipeline [25], which enables parameterization of standard amino acids. The pipeline cannot accommodate the hydroxyproline group because it does not belong to the standard amino acid set. In this instance, correct parameterization required the PRODRG program [26]. In the 1DZI complex, a fragment of trimeric GPOGPO**GFOGER**GPOGPOGPO 21-mer peptide has a high affinity sequence (**GFOGER**) designed to be recognized specifically by the $\alpha_2\beta_1$ integrin. In this sequence, the hydroxyproline group **O** has a fundamental role in the stabilization of the triple helical collagen [4], [5]. The relative dielectric constant is 78 for the solvent and 4 for the 1DZI complex, the ionic strength of the solution is 10 mM, and the temperature is 298 K. Initially, the electrostatic binding affinity between the collagen fragment and the $\alpha_2$-I domain was analyzed with no metal cation in the MIDAS site. It is meaningful to identify and investigate the electrostatics near the vicinity of a few essential amino acids that are potentially complementary. Initially, the collagen type-I and the $\alpha_2$-I domain are, respectively, the receptor ( R ) and ligand ( L ); later on, their roles were reversed.

In Fig. 1, certain essential amino acids are identified in the region known to be involved in the interaction of the residues, namely, the glutamic acid side chain of the GFOGER collagen motif and the metal ion in the MIDAS site.

**[Figure 1]**

**Figure 1.** Binding pocket region around the GFOGER motif in the 21-mer collagen strands. The GFOGER motif interacts with the $\alpha_2$-I domain. Ligands around the metal cation $Co^{2+}$ (red sphere) are $Ser^{153}$, $Ser^{155}$, $Thr^{221}$ of the $\alpha_2$-I domain, and two water molecules (not shown). The $\alpha_2$-I domain establishes weak ionic interactions with $Glu^{256}$ (3.16 Å) and two other $Asp^{151-254}$ residues in blue within 5Å to the $Co^{2+}$ are shown. Collagen glutamate $Glu^{11}$ forms a direct bond to the $Co^{2+}$ ion of the $\alpha_2$-I domain.

The relevant region involved in the electrostatic interaction corresponds to a divalent cation interacting with four amino acids in the binding pocket. These are $Glu^{11}$ that is bonded directly to the cation in the collagen chain, $Thr^{221}$, $Ser^{155}$, and $Ser^{153}$ in the $\alpha_2$-I domain, and two water molecules that complete the octahedral coordination. Three other negatively charged residues ($Asp^{151-254}$ and $Glu^{256}$) within 5 Å of the central $Co^{2+}$ atom surround the binding pocket. The corresponding interaction, desolvation, and residual potentials, computed on the surface of the $\alpha_2$-I domain, are mapped in Fig. 2. First, the electrostatic potential is extrapolated on the molecular surface of the $\alpha_2$-I integrin domain without a metal ion in the MIDAS site. Subsequently, the same extrapolation is obtained projecting the electrostatic potential on the integrin molecular surface with a $Co^{2+}$ cation in the MIDAS site.

**[Figure 2]**

**Figure 2**. A) Maps of electrostatic interaction plotted on solvent-accessible surfaces surrounding the MIDAS site of the $\alpha_2$-I domain. From left to right, the interaction potential $\Phi_R^i(\mathbf{x})$, the desolvation potential $\Phi_L^d(\mathbf{x})$, and the residual electrostatic potential $\Phi(\mathbf{x})$ computed on the ligand surface in absence of a metal cation. B) Maps of the electrostatic interaction on the ligand surface with $Co^{2+}$ in the MIDAS site. The surface charge distributions were calculated using the BEM code [17] with $I_s=10$ mM, T=298 K, relative solvent dielectric $\varepsilon_s=78$, and relative solvent dielectric $\varepsilon_m=4$.

In the absence of a metal ion, all the potential maps exhibit high negative values in the corresponding MIDAS site. The interaction potential, generated by the collagen charges alone, has a negative peak localized around the prominent glutamate group labeled by $Glu^{11}$ in the 1DZI pdb file described in [7]. The desolvation potential generated by the integrin domain charges is due to the negative aspartate and glutamate groups labeled by $Asp^{151-254}$ $Glu^{256}$. The comparable negative values in both the interaction and desolvation potentials lead to high negative values of the residual potential in the MIDAS region and, therefore, to low electrostatic affinities. This was supported by the fact that the Pearson correlation coefficient is a positive 0.19. As pointed out in [20], the analysis based on the residual potential is intrinsically asymmetric because, in general, the receptor and the ligand cannot be mutually complementary. Moreover, the degree of complementarity is affected by the contribution of ligand and receptor functions. In this specific case, the same conclusions can be drawn when the collagen molecule is the ligand, and the $\alpha_2$-I integrin domain is the receptor, by computing the difference potentials on its surface. In Fig. 1, the complementarity principle is almost satisfied, because the interaction and the desolvation electrostatic potentials have nearly opposite signs relative to the electrostatic desolvation and interaction potentials. This is a consequence of the complementarity between the collagen and $\alpha_2$-I integrin domains in the presence of a stabilizing $Co^{2+}$ cation. Thus, the reaction potential is approximately zero on most of the integrin domain surface. The main result of this study is to confirm that the binding mechanism is fundamentally influenced by the $Co^{2+}$, $Mg^{2+}$, and $Mn^{2+}$ cations, which are known to support integrin-collagen adhesion *in vivo* [5]. The presence of $Ca^{2+}$ leads to a more complex behavior and thus requires a more detailed study. The simulations performed for $\alpha_2$-I integrin domain and collagen bonded to $Mg^{2+}$ and $Mn^{2+}$ cations show a complementary region for each screened cation localized in the MIDAS site. In all cases, the results obtained for the ligand-receptor bound complex with the residual potential give a clear indication of the stabilizing effects produced by a specific cation. The electrostatic complementarity between the triple helical collagen fragment and the $\alpha_2$-I domain was also quantified with the Pearson correlation coefficient. The results plotted in Fig. 3 provide evidence that the Pearson coefficient is cation dependent, as shown by the residual potential projected onto the molecular surface of the $\alpha_2$-I domain. The correlation in Fig. 2 drops in the presence of a divalent cation, and reflects the fact that, in the 21-mer collagen, most regions with positive values of the desolvation potential match regions with opposite values of the interaction potential. However, the bound complex does not reach perfect complementarity, therefore the correlation does not assume the value -1. In fact, the residual potential maintains positive values near the binding pocket.

**[Figure 3]**

**Figure 3**. Plot of the correlation coefficient as function of ionic radius computed for $I_s=10$ mM. The plot takes into account two cases. The first one (red line) is obtained using the complete structure to find the Pearson coefficient. The second one (green line) is obtained using the reduced structure, with the fundamental four amino acids in the binding pocket. The Glu[11] in the collagen chain and Thr[221], Ser[155], and Ser[153] in the $\alpha_2$-I domain.

Figure 3 also shows the Pearson coefficient that is obtained when taking into account only the charges of the fundamental amino acids involved in the local electrostatic interaction. The results agree with the previous ones. This evidence is supported by the value of the Pearson coefficient due to the presence of the fundamental binding pocket. The Pearson coefficient is also higher compared with the full charge simulation for all metal ions simulated independently from the cation located in the MIDAS site.

**Association rates of the $\alpha_2$-I domain and 21-mer collagen fragment with and without $Mg^{2+}$, $Co^{2+}$ and $Mn^{2+}$**

The $k_{on}$ rate constants shown in [5] suggest that collagen and the $\alpha_2$-I domain find their specific binding site through the allowed reciprocal orientation and the electrostatic interaction. The "*Browndye simulation code*" [27] is employed for this purpose, treating proteins as rigid bodies. Effective charges are used to reproduce pre-computed electrostatic potentials. The influence of these potentials on diffusional motion is determined from the standard "*Ermak and McCammon*" algorithm [28]. Association rates are computed for $I_s$=10 mM [5]. An adaptive time step with a minimum value of 1 ps is chosen, and trajectories are propagated until the transient complex is obtained. The inset in Fig. 4 plots the $k_{on}$ rates in the open state conformation, where the binding is strongly supported by the electrophilic metal cations $Co^{2+}$, $Mg^{2+}$, and $Mn^{2+}$ that are able to promote the formation of a octahedral coordination at the MIDAS site. A reference is the case without a metal ion. These results were obtained by running $5 \cdot 10^4$ Brownian trajectories that take into account only the effect of the electrostatic interaction. When Glu[11] and a cation are less than 6 Å apart (the criterion to define the formation of an encounter complex with two contacts), the simulations produce higher values of $k_{on}$ relative to the experimental values. The second group of simulations is performed with the same criteria of the previous ones, but with additional desolvation and hydrodynamic terms. In this case, the evaluation of the $k_{on}$ rate requires more time-consuming simulations for statistical significance. Thus, the simulation ran over $10^6$ trajectories for the $\alpha_2$-I-collagen complex. The $k_{on}$ values range over $3 \cdot 10^4$–$2 \cdot 10^5$ and are compatible with the experimental data in [5]. In Fig. 4, the average $k_{on}$ rates are ordered $k_{on}^{Co^{2+}} > k_{on}^{Mg^{2+}} > k_{on}^{Mn^{2+}}$ between 6 Å and 8 Å. All the results are also analyzed in terms of statistical inferences based on the overlap of confidence intervals (CIs). The criterion here adopted is that 83% of CIs of two averages do not overlap to assess whether or not the averages are significantly different from each other at the $p \leq .05$ level [29]. Figure 4 shows that the average $k_{on}^{Mn^{2+}}$ is statistically distinguishable from the average values of $k_{on}^{Co^{2+}}$ and $k_{on}^{Mg^{2+}}$. However, when comparing $k_{on}^{Co^{2+}}$ and $k_{on}^{Mg^{2+}}$, it is not possible to infer any statistical difference.

**[Figure 4]**

**Figure 5**. $k_{on}$ of the $\alpha_2$-I integrin domain and 21-mer collagen peptide in the open state conformation. The inset was obtained taking into account only the electrostatic interaction computed with the APBS code and using 10 nested grids, assuming a salt concentration of 10 mM, a relative solvent dielectric equal to 4, a relative water dielectric equal to 78.5, and a temperature fixed at 298 K. The binding events were defined according to a criterion of three contacts of possible hydrogen-bonding pairs at a separation < 6 Å, and $5 \cdot 10^4$ trajectories. The plot also takes into account the hydrodynamic and desolation force fields obtained with the same value regarding the chemical and the physical parameters. The red, green, and blue lines depict $k_{on}$ for $Mg^{2+}$, $Co^{2+}$, and $Mn^{2+}$, respectively, for $10^6$ trajectories. The data are reported according to the 83% confidence intervals (CIs) criterion.

**Conclusions**

This work presents realistic simulations of binding events relevant to cell adhesion processes. The presence of cations is a key factor affecting the electrostatic binding interactions between an $\alpha_2$-I integrin domain and a fragment of collagen peptide. This effect is particularly well demonstrated by the presence of the divalent cation $Co^{2+}$. As depicted in Fig. 2, the numerical simulation reveals the central role of electrostatic correlation when complementarity is assumed. The electrostatic complementarity is determined by the correlation coefficient and is a monotonically decreasing function of the ionic radius. The interaction potential $\Phi_r^i(\mathbf{x})$ due to the electric charge of the $\alpha_2$-I integrin domain, and the desolvation potential $\Phi_1^d(\mathbf{x})$ due to the charges of the triple helical collagen, are more anti-correlated if a metal cation is located in the MIDAS site. Therefore, a larger metal cation radius creates greater electrostatic affinity in the bound state of the $\alpha_2$-I integrin-collagen complex. Thus, divalent cations probably bind collagen to the integrin domain by bridging both molecules. The underlying assumptions of a Brownian dynamics approach, as well as the implicit water continuum, have some limitations; however, it is possible to statistically distinguish $Mn^{2+}$ from $Co^{2+}$ and $Mg^{2+}$ according to different association rates in the formation of the encounter complex. Finally, as the ionic radius of the cation increases, both the association rates and the hydrogen-bonding lengths of the pairs decrease. Finally, the $k_{on}$ values are compatible with the experimental data in [5]. Therefore, the assumption of binding kinetics in the diffusion-limited regime could faithfully describe metal-ion-driven protein binding. There are two drawbacks of this method. The first is that divalent cations have very specific interactions with water that are not completely described in the Poisson Boltzmann equation [31]. The second is that association rates depend upon the criteria selected to identify a successful binding.

**Competing interests**

The authors declare that they have no competing interests.

**Acknowledgment**

**References**

1. Dickeson SK, Santoro SA: **Ligand recognition by the I domain-containing integrins**. *Cell Mol Life Sci* 1998, **54**:556–566.

2. White DJ, Puranen S, Johnson MS, Heino J: **The collagen receptor subfamily of the integrins.** *Int J Biochem Cell Biol* 2004, **36**:1405–10.

3. Tuckwell D, Calderwood DA, Green LJ, Humphries MJ: **Integrin alpha 2 I-domain is a binding site for collagens.** *J Cell Sci* 1995, **108 ( Pt 4**:1629–37.

4. Takada Y, Wayner EA, Carter WG, Hemler ME: **Extracellular matrix receptors, ECMRII and ECMRI, for collagen and fibronectin correspond to VLA-2 and VLA-3 in the VLA family of heterodimers.** *J Cell Biochem* 1988, **37**:385–93.

5. Siebert H-C, Burg-Roderfeld M, Eckert T, Stötzel S, Kirch U, Diercks T, Humphries MJ, Frank M, Wechselberger R, Tajkhorshid E, Oesser S: **Interaction of the α2A domain of integrin with small collagen fragments.** *Protein cell* 2010, **1**:393–405.

6. Obsil T, Hofbauerová K, Amler E, Teisinger J: **Different cation binding to the I domains of alpha1 and alpha2 integrins: implication of the binding site structure.** *FEBS Lett* 1999, **457**:311–5.

7. Emsley J, Knight CG, Farndale RW, Barnes MJ: **Structure of the Integrin α2β1-binding Collagen Peptide.** *J Mol Biol* 2004, **335**:1019–1028.

8. Xiong JP, Stehle T, Diefenbach B, Zhang R, Dunker R, Scott DL, Joachimiak A, Goodman SL, Arnaout MA: **Crystal structure of the extracellular segment of integrin alpha Vbeta3.** *Science 2001*, **294**:339–45.

9. Kamata T: **Interaction between Collagen and the alpha 2 I-domain of Integrin alpha 2beta 1. Critical role of conserved residues in the metal ion-dependent adhesion site (midas) region.** *J Biol Chem 1999*, **274**:32108–32111.

10. Emsley J, King SL, Bergelson JM, Liddington RC: **Crystal Structure of the I Domain from Integrin α2β1.** *J Biol Chem 1997*, **272**:28512–28517.

11. Emsley J, Knight CG, Farndale RW, Barnes MJ, Liddington RC: **Structural basis of collagen recognition by integrin alpha2beta1.** *Cell* 2000, **101**:47–56.

12. Gilson MK: **Theory of electrostatic interactions in macromolecules.** *Curr Opin Struct Biol* 1995, **5**:216–23.

13. Lu BZ, Zhou YC, Holst MJ, McCammon JA: **Recent progress in numerical methods for the Poisson-Boltzmann equation in biophysical applications.** *Commun Comput Phys* 2008, **3**:973–1009.

14. Baker NA, Sept D, Joseph S, Holst MJ, McCammon JA: **Electrostatics of nanosystems: application to microtubules and the ribosome.** *Proc Natl Acad Sci USA* 2001, **98**:10037–10041.

15. Honig B, Nicholls A: **Classical electrostatics in biology and chemistry.** *Science* 1995, **268**:1144-9.

16. Koehl P, Delarue M: **AQUASOL: An efficient solver for the dipolar Poisson-Boltzmann-Langevin equation.** *J Chem Phys* 2010, **132**:064101.

17. Manzin A, Bottauscio O, Ansalone DP: **Application of the thin-shell formulation to the numerical modeling of Stern layer in biomolecular electrostatics.** *J Comput Chem* 2011, **32**:3105–3113.

18. Manzin A, Ansalone DP, Bottauscio O: **Numerical Modeling of Biomolecular Electrostatic Properties by the Element-Free Galerkin Method**. *In IEEE Trans Magn*. Volume 47. IEEE; 2011:1382–1385.

19. Lee L-P, Tidor B: **Optimization of electrostatic binding free energy**. *J Med Phys* 1997, **106**:8681–8690.

20. Kangas E, Tidor B: **Optimizing electrostatic affinity in ligand–receptor binding: Theory, computation, and ligand properties**. *J Chem Phys* 1998, **109**:7522–7545.

21. Lee L-P, Tidor B: **Optimization of binding electrostatics: charge complementarity in the barnase-barstar protein complex.** *Protein Sci* 2001, **10**:362–377.

22. Joughin BA, Green DF, Tidor B: **Action-at-a-distance interactions enhance protein binding affinity.** *Protein Sci* 2005, **14**:1363–9.

23. Bongrand P: **Ligand-receptor interactions**. *Reports Prog Phys* 1999, **62**:921–968.

24. Alsallaq R, Zhou H-X: **Electrostatic rate enhancement and transient complex of protein-protein association.** *Proteins* 2008, **71**:320–35.

25. Dolinsky TJ, Czodrowski P, Li H, Nielsen JE, Jensen JH, Klebe G, Baker NA: **PDB2PQR: expanding and upgrading automated preparation of biomolecular structures for molecular simulations.** *Nucleic Acids Res* 2007, **35**(Web Server issue):W522–5.

26. Schüttelkopf AW, van Aalten DMF: **PRODRG: a tool for high-throughput crystallography of protein-ligand complexes.** *Acta Crystallogr D Biol Crystallogr* 2004, **60**(Pt 8):1355–63.

27. Huber G a, McCammon JA: **Browndye: A Software Package for Brownian Dynamics.** *Comput Phys Commun* 2010, **181**:1896–1905.

28. Ermak DL, McCammon JA: **Brownian dynamics with hydrodynamic interactions**. *J Chem Phys* 1978, **69**:1352.

29. Austin PC, Hux JE: **A brief note on overlapping confidence intervals**. *J Vasc Surg* 2002, **36**:194–195.

30. Anthony PC, Sim AYL, Chu VB, Doniach S, Block SM, Herschlag D: **Electrostatics of nucleic acid folding under conformational constraint.** *J Am Chem Soc* 2012, **134**:4607–14.31

31. Yufeng L, Esmael H, Tobin RS, Karl FF, Haipeng G: **A Novel Implicit Solvent Model for Simulating the Molecular Dynamics of RNA**, *Biophysical Journal*, 105:1248-57.

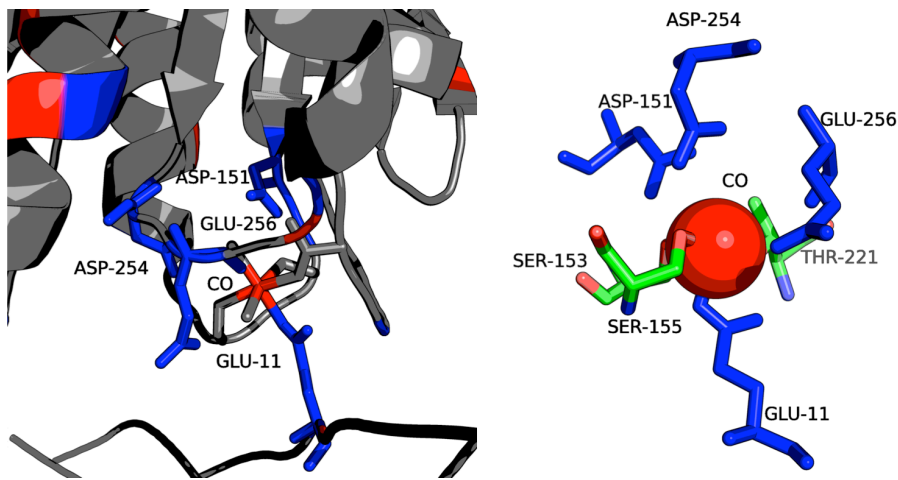

**Figure 1.** Binding pocket region around the GFOGER motif in the 21-mer collagen strands. The GFOGER motif interacts with the $\alpha_2$-I domain. Ligands around the metal cation $Co^{2+}$ (red sphere) are $Ser^{153}$, $Ser^{155}$, $Thr^{221}$ of the $\alpha_2$-I domain, and two water molecules (not shown). The $\alpha_2$-I domain establishes weak ionic interactions with $Glu^{256}$ (3.16 Å) and two other $Asp^{151-254}$ residues in blue within 5Å to the $Co^{2+}$ are shown. Collagen glutamate $Glu^{11}$ forms a direct bond to the $Co^{2+}$ ion of the $\alpha_2$-I domain.

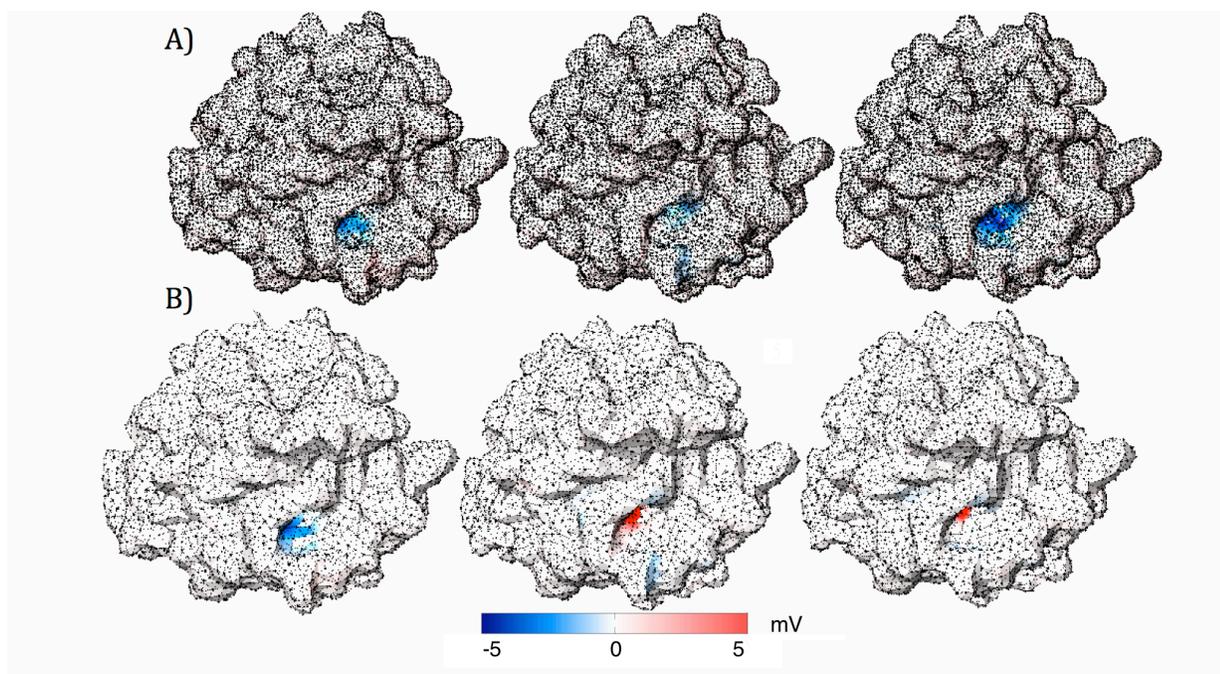

**Figure 2**. A) Maps of electrostatic interaction plotted on solvent-accessible surfaces surrounding the MIDAS site of the $\alpha_2$-I domain. From left to right, the interaction potential $\Phi_R^i(\mathbf{x})$, the desolvation potential $\Phi_L^d(\mathbf{x})$, and the residual electrostatic potential $\Phi(\mathbf{x})$ computed on the ligand surface in absence of a metal cation. B) Maps of the electrostatic interaction on the ligand surface with $Co^{2+}$ in the MIDAS site. The surface charge distributions were calculated using the BEM code [17] with $I_s$=10 mM, T=298 K, relative solvent dielectric $\varepsilon_s$=78, and relative solvent dielectric $\varepsilon_m$=4.

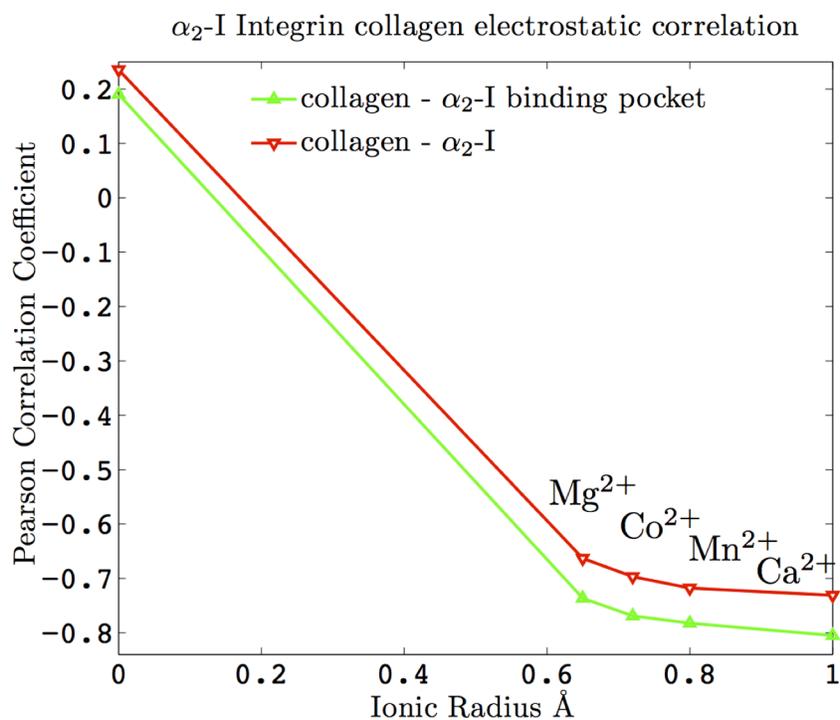

**Figure 3**. Plot of the correlation coefficient as function of ionic radius computed for $I_s$=10 mM. The plot takes into account two cases. The first one (red line) is obtained using the complete structure to find the Pearson coefficient. The second one (green line) is obtained using the reduced structure, with the fundamental four amino acids in the binding pocket. The $Glu^{11}$ in the collagen chain and $Thr^{221}$, $Ser^{155}$, and $Ser^{153}$ in the $\alpha_2$-I domain.

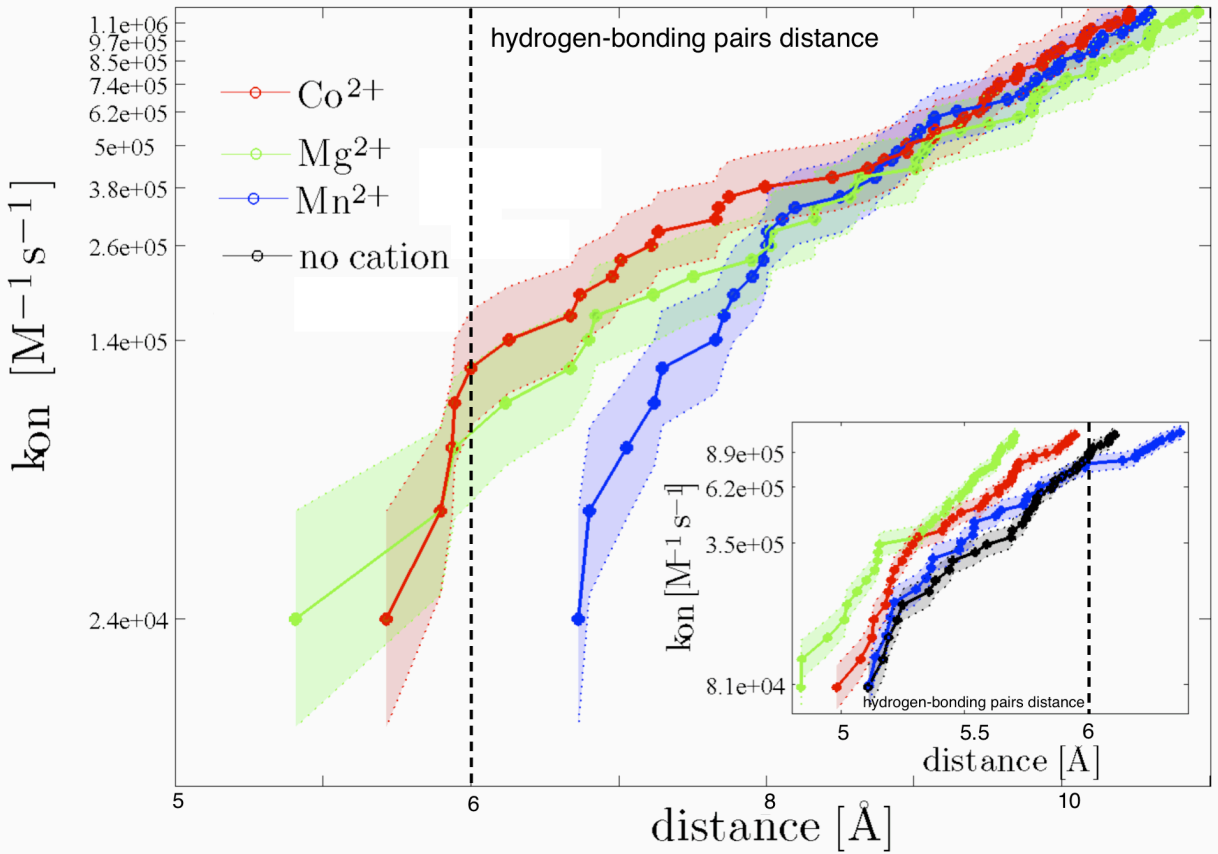

**Figure 6**. $k_{on}$ of the $\alpha_2$-I integrin domain and 21-mer collagen peptide in the open state conformation. The inset was obtained taking into account only the electrostatic interaction computed with the APBS code and using 10 nested grids, assuming a salt concentration of 10 mM, a relative solvent dielectric equal to 4, a relative water dielectric equal to 78.5, and a temperature fixed at 298 K. The binding events were defined according to a criterion of three contacts of possible hydrogen-bonding pairs at a separation < 6 Å, and $5 \cdot 10^4$ trajectories. The plot also takes into account the hydrodynamic and desolvation force fields obtained with the same value regarding the chemical and the physical parameters. The red, green, and blue lines depict $k_{on}$ for $Mg^{2+}$, $Co^{2+}$, and $Mn^{2+}$, respectively, for $10^6$ trajectories. The data are reported according to the 83% confidence intervals (CIs) criterion.